\documentclass[aps,pra,twocolumn,showpacs,amsmath,amssymb,10pt]{revtex4-1}
\usepackage{graphicx}
\usepackage{bm}
\usepackage{slashed}

\begin{document}
\title{Universality of comb structures in strong-field QED}
\author{K. Krajewska}
\email[E-mail address:\;]{Katarzyna.Krajewska@fuw.edu.pl}
\author{J. Z. Kami\'nski}
\affiliation{Institute of Theoretical Physics, Faculty of Physics, University of Warsaw, Ho\.{z}a 69,
00-681 Warszawa, Poland}
\date{\today}
\begin{abstract}
We demonstrate a new mechanism for coherent comb generation in both radiation and matter domains, which is
by using strong-field quantum electrodynamics processes. Specifically, we demonstrate this for nonlinear Compton scattering
which has the potential to generate frequency combs extending towards the $\gamma$-ray spectral region as well as combs of ultra-relativistic 
electrons extending towards the MeV regime. Moreover, we show that coherent energy combs of positrons (or electrons) emerge in 
laser-induced pair creation processes such as the Breit-Wheeler process, proving the universality of the proposed mechanism.
\end{abstract}
\maketitle

Historically, strong-field quantum electrodynamics (QED) is the oldest strong-field-physics area,
as investigations of fundamental QED processes in laser fields began in the 1960s~\cite{QEDold} 
(for recent reviews see,~\cite{Mourou,Ehlotzky,HeidelbergReview}). 
At that time, the problem was of purely theoretical interest and the main focus was on a
qualitative understanding of effects imposed by intense and mostly monochromatic electromagnetic 
radiation. With the rapid development of high-power laser technology, that we have encountered over
the last decades, the laser radiation in the visible spectrum with intensities as high as $10^{22}$ W/cm$^2$
can be produced in the laboratory~\cite{Mourou,Yanovsky}. At the same time, the synthesis of laser pulses with well-controlled
properties, such as the carrier-envelope phase or the envelope shape, is currently feasible. Due to those recent
advances, it became urgent to quantitatively investigate more subtle effects of the laser-field interaction 
with matter than was originally pursued.

For now, the explorations of laser field interaction with matter in the area of 
strong-field QED are mostly carried out theoretically, paving the ground for future experimental 
investigations in the rapidly progressing areas of relativistic optics and information theory. To name 
only a few of the related studies let us mention: the vacuum polarization induced double-slit Young-type interference~\cite{YoungKing}, 
the generation of entangled states of high-energy photons by means of the double or triple Compton process~\cite{entangledphotons}, 
the phase control of laser-induced or laser-modified fundamental QED processes~\cite{KKphase,KKcompton,KKbw}, 
the light diffraction and spin dynamics induced by a strong standing electromagnetic wave 
(related to the Kapitza-Dirac effect)~\cite{kapica}, and spin effects in the electron-positron pair creation~\cite{CM13}. 
Surprisingly, the strong-field QED vacuum instability and pair production have found their analogues in solid-state physics~\cite{solid}.

In this Letter we propose a novel method for generating $\gamma$-ray frequency combs, which is by 
nonlinear Compton scattering. In addition, we demonstrate the formation of coherent particle combs,
which can be achieved not only via nonlinear Compton scattering but via other strong-field QED 
processes as well. In this sense, the universality of comb structures in strong-field QED is emerging.
As an illustration, we demonstrate the generation of positron energy combs via the nonlinear Breit-Wheeler process.

The optical frequency comb has become an indispensable tool for high precision spectroscopy \cite{FreqComb}.
Also experiments in the field of ultrafast physics rely on the frequency comb technique to generate precisely controlled attosecond 
optical pulses by means of the high-order harmonic generation (see, e.g., Ref.~\cite{HHGComb}). However, in order 
to generate even shorter laser pulses or to apply this technique in investigations of nuclear structure, combs 
of frequencies of the order of MeV are necessary. It seems that it may not be possible to achieve such photon energies 
by high-order harmonic generation (the generation of soft x-ray harmonic combs by relativistic electron spikes 
has been realized recently \cite{Pirozhkov12}, but not for such high frequencies). In this context, we study
the possibility of the generation of Compton-scattering-based frequency combs.

One of the most intensively studied fundamental QED processes is the laser-induced or laser-modified nonlinear 
Compton scattering~\cite{KKcompton,ComptonOthers}. It allows for the generation of intense electromagnetic radiation 
in the photon energy domain comparable to, or even exceeding, the electron rest energy. The differential energy 
distribution of Compton photons emitted in this process adopts the form (see, Eq. (51) in Ref.~\cite{KKcompton}, 
where all symbols and physical units used in this Letter are defined)
\begin{equation}
\frac{d^3E_{\mathrm{C}}(\bm{K}\sigma;\bm{p}_{\mathrm{i}}\lambda_{\mathrm{i}})}{d\omega_{\bm{K}}d^2\Omega_{\bm{K}}}=\sum_{\lambda_{\mathrm{f}}}
\frac{\alpha(m_{\mathrm{e}}c)^2k^0(K^0)^2}{(2\pi)^2p_{\mathrm{i}}^0(k\cdot p_{\mathrm{f}})}|\mathcal{A}_{\mathrm{C}}|^2,
\label{compton1}
\end{equation} 
with
\begin{equation}
\mathcal{A}_{\mathrm{C}}=\sum_N D_N\frac{1-e^{-2\pi iP_N^0/k^0}}{iP_N^0},
\label{compton2}
\end{equation}
and with the components $D_N$ defined by Eqs. (23) and (44) in~\cite{KKcompton}. Let us also introduce 
the phase of the Compton probability amplitude $\Phi_{\mathrm{C}}=\arg(\mathcal{A}_{\mathrm{C}})$ such that
$-\pi<\arg(\mathcal{A}_{\mathrm{C}})\leqslant\pi$. We define a linearly polarized laser pulse which propagates 
in the $z$-direction such that the corresponding vector potential equals $\bm{A}(k\cdot x)=A_0\bm{e}_x f(k\cdot x)$ for 
$0\leqslant k\cdot x\leqslant 2\pi$, and vanishes otherwise. The four-vector $k$ equals $(\omega/c)(1,\bm{e}_z)$, whereas 
the frequency $\omega=2\pi/T_{\mathrm{p}}$ is determined by the duration of the laser pulse, $T_{\mathrm{p}}$. 
The shape function, $f(k\cdot x)$, is an arbitrary differentiable real function of its argument, so that the 
electric field component is $\bm{\mathcal{E}}(k\cdot x)=\mathcal{E}_0\bm{e}_x f'(k\cdot x)$, where \textit{'prime'} means 
the derivative with respect to the argument, and $\mathcal{E}_0=-A_0\omega$.

\begin{figure}
\includegraphics[width=6cm]{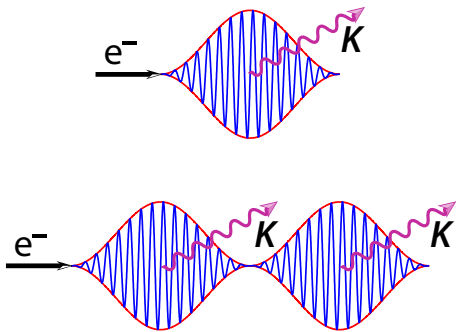}
\caption{(Color online) The Young-type experiment related to nonlinear Compton scattering by short and modulated 
laser pulses. Electrons in a head-on collision with the laser pulse, in general consisting of $N_{\mathrm{rep}}$ 
sub-pulses, can emit from each of them a Compton photon $K$. The interference of QED probability amplitudes leads 
to the coherent enhancement of emitted radiation. This happens at equally-spaced frequencies, resulting in the 
formation of frequency combs in the spectrum of Compton photons. \label{youngmgtwo}}
\end{figure}

As it is known from previous studies~\cite{KKcompton,ComptonOthers}, the Compton photons energy spectrum 
for intense and short driving pulses consists of broad incoherent peaks in the sense that for the peak-related 
frequencies the QED probability amplitudes have uncontrollable phases $\Phi_{\mathrm{C}}$. In order to improve this,
we consider a strong-field QED analogue of the Young interference experiment~\cite{Young1804}; namely, instead 
of a single pulse, we apply a modulated pulse. More specifically, we consider 
a pulse composed of $N_{\mathrm{rep}}$ identical sub-pulses, which is illustrated in Fig. \ref{youngmgtwo} 
(lower graphic) for $N_{\mathrm{rep}}=2$. Each sub-pulse acts as a single slit in the Young experiment, 
from which a Compton photon is emitted with a particular probability amplitude. When adding these amplitudes one 
can expect that probabilities for emission of Compton photons are enhanced or suppressed for some particular 
frequencies. As we demonstrate below for a given direction of observation, this leads to the formation of 
frequency combs. In order to illustrate this effect we choose the shape function 
$f(\phi)=N_{\mathrm{rep}}N_{\mathrm{osc}}f_0(\phi)$ where
\begin{equation}
f'_0(\phi)=N_{\mathrm{A}}\sin^2\Bigr(N_{\mathrm{rep}}\frac{\phi}{2}\Bigl)\sin(N_{\mathrm{rep}}N_{\mathrm{osc}}\phi), 
\label{compton3}
\end{equation}
for $0\leqslant \phi \leqslant 2\pi$ and 0 otherwise. Such a laser pulse consists of $N_{\mathrm{rep}}$
sub-pulses, each including $N_{\mathrm{osc}}$ field oscillations (this particular shape function is
illustrated in Fig. \ref{youngmgtwo} for $N_{\mathrm{osc}}=16$, and $N_{\mathrm{rep}}=1$ or 2). The normalization 
constant $N_{\mathrm{A}}$ is chosen such that
\begin{equation}
\frac{1}{2\pi}\int_0^{2\pi} d\phi [f'(\phi)]^2=\frac{1}{2},
\label{compton4}
\end{equation}
and the carrier frequency equals $\omega_{\mathrm{L}}=N_{\mathrm{rep}}N_{\mathrm{osc}}\omega$. As it was argued in 
Ref.~\cite{KKphase}, imposing such a condition [Eq.~\eqref{compton4}] guarantees that the average laser field intensity contained in the
pulse is fixed, no matter how many sub-pulses or cycles in each sub-pulse is included. For such definitions the 
averaged intensity of the laser field (given in W/cm$^2$) equals
$I=a\omega_{\mathrm{L}}^2\mu^2$,
where the dimensionless and relativistically invariant parameter $\mu$ is $|eA_0|/m_{\mathrm{e}}c$, the carrier frequency
$\omega_{\mathrm{L}}$ is in relativistic units of energy, and $a=2.3\times 10^{29}$.

\begin{figure}
\includegraphics[width=8cm]{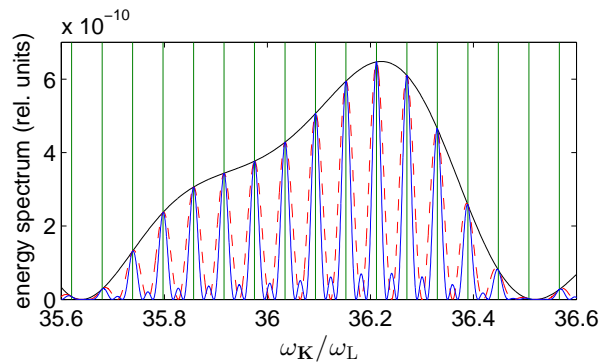}
\caption{(Color online) The energy spectrum (in rel. units), Eq.~\eqref{compton1}, summed over the final photon polarizations 
$\sigma$ and averaged over the initial electron spins $\lambda_{\mathrm{i}}$. The reference frame is chosen such that initial 
electrons are at rest, i.e., $\bm{p}_{\mathrm{i}}=0$. In addition, $\omega_{\mathrm{L}}=4.15\times 10^{-4}m_{\mathrm{e}}c^2$, $\mu=1$, 
$N_{\mathrm{osc}}=16$, and the direction of emission of Compton photons is given by the polar and azimuth angles, 
$\theta_{\bm{K}}=0.2\pi$ and $\varphi_{\bm{K}}=0$. The broad peak under the thin continuous black line corresponds 
to $N_{\mathrm{rep}}=1$, the red dashed line is for $N_{\mathrm{rep}}=2$, whereas the continuous blue one is for $N_{\mathrm{rep}}=3$. 
With increasing $N_{\mathrm{rep}}$, the main peaks become narrower, and in addition extra $(N_{\mathrm{rep}}-2)$ small
peaks appear between them, which is the typical interference or diffraction effect. The energy spectrum for $N_{\mathrm{rep}}$ 
sub-pulses is divided by $N_{\mathrm{rep}}^2$ in order to show that the spectrum for $N_{\mathrm{rep}}$ sub-pulses at the 
peak-frequencies is coherently enhanced. The vertical thin green lines are equally spaced, which shows that Compton photons 
form the frequency comb.
\label{comb1_20130301v}}
\end{figure}

\begin{figure}
\includegraphics[width=8cm]{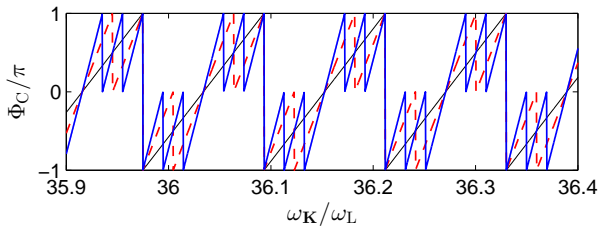}
\caption{(Color online) The phase $\Phi_{\mathrm{C}}$ of the QED probability amplitude [Eq.~\eqref{compton2}] 
as the function of emitted Compton photon energy for the same parameters as in Fig.~\ref{comb1_20130301v}. It is 
equal to 0 modulo $\pi$ for discrete values of $\omega_{\bm{K}}=\omega_{\ell}$ (with integer $\ell$) such that 
$\omega_{\ell}/\omega_{\mathrm{L}}=35.9154+\ell\cdot\Delta_{\mathrm{comb}}$ with $\Delta_{\mathrm{comb}}=0.0444$. 
Exactly for these frequencies the energy spectrum presented in Fig. \ref{comb1_20130301v} shows the main peaks.
Similar to Fig.~\ref{comb1_20130301v}, the solid and thin black line relates to $N_{\rm rep}=1$, the red dashed line is for
$N_{\rm rep}=2$, whereas the blue solid line is for $N_{\rm rep}=3$.
\label{combphase20130317f}}
\end{figure}

Let us consider the electron beam parameters reported in Ref.~\cite{Huang98}, i.e., the 35 MeV electron beam with 
the energy spread 0.1\%. For the laser pulse we choose the Ti-Sapphire laser with $\omega_{\mathrm{L}}=1.548$ eV. 
In Fig. \ref{comb1_20130301v} we present the differential energy distribution in the reference frame in which 
initial electrons are at rest. As mentioned above, for a nonperturbative laser pulse, this distribution consists 
of broad peaks, one of which is represented in this figure by a thin black line. Note that it scans a frequency range 
whose width is approximately equal to the carrier frequency $\omega_{\mathrm{L}}$. Moreover, the phase $\Phi_{\mathrm{C}}$ 
of the QED probability amplitude takes all values from the interval $[-\pi,\pi]$ (see, Fig. \ref{combphase20130317f}) 
which indicates that the emitted radiation is non-coherent, even though the driving pulse parameters are precisely controlled.

Let us investigate the possibility of generating the coherent Compton radiation. For this purpose, we consider the incident 
pulse which consists of a finite sequence of $N_{\mathrm{rep}}$ sub-pulses, as it is illustrated in Fig. \ref{youngmgtwo}. 
In the case when $N_{\mathrm{rep}}>1$, we observe the emergence of the regular comb structure in which peak positions are 
independent of $N_{\mathrm{rep}}$. It follows from this figure that the resolved comb frequencies are equally spaced.
At those peak frequencies, the energy distribution scales like $N_{\mathrm{rep}}^2$, whereas widths of individual peaks become increasingly narrower, with increasing 
$N_{\mathrm{rep}}$. In order to demonstrate that the comb consists of coherent radiation, in Fig. \ref{combphase20130317f} 
we plot the phase $\Phi_{\mathrm{C}}$ which at the comb frequencies adopts the same value modulo $\pi$. This phase
coherence makes it promising to use the proposed Young-type mechanism for generation of zepto- or even yoctosecond laser pulses.

\begin{figure}
\includegraphics[width=8cm]{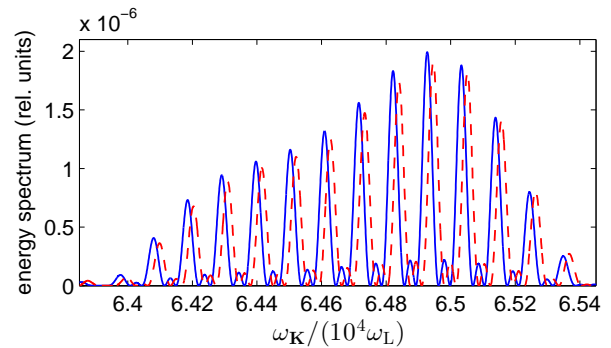}
\caption{(Color online) The same as in Fig.~\ref{comb1_20130301v} but in the laboratory frame in which the head-on collision
of an electron and a laser beam takes place. We choose the parameters such that $\omega_{\mathrm{L}}=1.548\, \mathrm{eV}$, 
$N_{\mathrm{rep}}=3$, $\theta_{\bm{K}}=0.9857\pi$, and $\varphi_{\bm{K}}=0$. The results are compared for the case when the initial 
electron momentum is $|\bm{p}_{\mathrm{i}}|=68.493 m_{\mathrm{e}}c$ (solid line) and $|\bm{p}_{\mathrm{i}}|=68.563 m_{\mathrm{e}}c$ 
(dashed line), i.e., they differ by roughly 0.1\%. This corresponds to the electron beam of energy 35 MeV with the energy 
spread roughly 0.1\% (parameters as given in Ref.~\cite{Huang98}).
\label{comba_20130304}}
\end{figure}

In light of these results, an important question arises: How sensitive is the observed comb structure with respect to a change 
of the electron initial energy? To answer this question, in Fig. \ref{comba_20130304} we present the same comb structure as 
in Fig. \ref{comb1_20130301v}, but in the laboratory frame of reference for $N_{\mathrm{rep}}=3$ and for two electron initial 
energies which differ from each other by 0.1\%~\cite{Huang98}. As we see, when changing the initial electron energy, 
the comb structure shifts a bit but nevertheless, after averaging over the electron initial energy distribution, the comb peaks 
will be clearly resolved. This shows that the coherent comb structures  for the Compton process can be observed 
experimentally with currently available laser and electron beam parameters. Let us also note that performing the 
Young-type experiment with respect to nonlinear Compton scattering, which is by using a modulated pulse shown in Fig.~\ref{youngmgtwo}
for $N_{\rm rep}=2$, one could produce very energetic radiation combs; for the parameters employed in Fig. \ref{comb1_20130301v}, 
such a structure reaches the 100 keV regime. The same concerns the MeV regime, although the respective comb structures 
are smaller in magnitude.

\begin{figure}
\includegraphics[width=8cm]{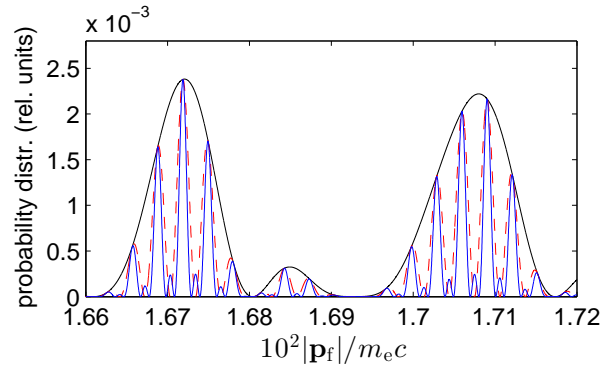}
\caption{(Color online) The differential probability distribution (in rel. units) of the electron scattered via the Compton process
[Eq.~\eqref{compton5}]. The laser beam and the initial electron parameters are the same as in Fig. \ref{comb1_20130301v}, but the final 
electron propagation direction is fixed at the angles $\theta_{\mathrm{f}}=0.2\pi$ and $\varphi_{\mathrm{f}}=0$. Just like for the Compton photons, 
for the scattered electrons the coherent comb structure is formed (the corresponding probability distributions are divided by $N_{\mathrm{rep}}^2$).
\label{boost20130330}}
\end{figure}
\begin{figure}
\includegraphics[width=8cm]{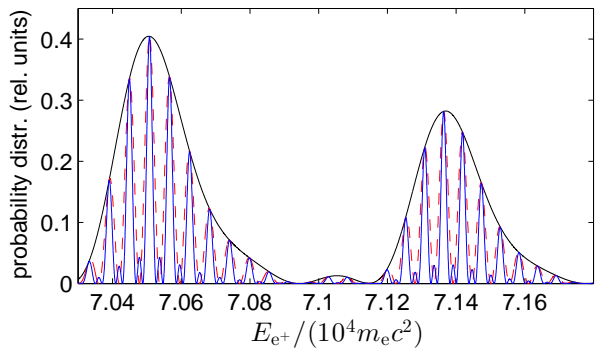}
\caption{(Color online) The positron differential probability distribution (in rel. units) for the Breit-Wheeler pair creation process 
calculated in the laboratory frame of reference. The laser pulse characterized by parameters $\omega_{\mathrm{L}}=1.548\,\mathrm{eV}$, $\mu=1$, and $N_{\mathrm{osc}}=16$
collides with a counterpropagating $\gamma$-photon of energy $10^5m_{\mathrm{e}}c^2$ and a linear polarization along the $x$-axis. The positrons are created 
in the direction such that $\theta_{\mathrm{e}^+}=0.99999\pi$ and $\varphi_{\mathrm{e}^+}=0$. The results have been summed up with respect to the
spin degrees of freedom of produced particles. As in the previous cases, the coherent
comb structure for created positrons (or electrons) is formed (the corresponding probability distributions are divided by $N_{\mathrm{rep}}^2$).
\label{bwcomb20130322par}}
\end{figure}

It is worth noting that similar structures are also observed in the Thompson radiation spectrum, though for smaller frequency 
ranges, because of the domain of applicability of the classical theory. Analysis of the relativistic Newton-Lorentz equation 
shows that the pulses used in this paper do not accelerate or decelerate electrons. Note that this is in contrast to the Compton process 
in which scattered electrons change their energy. Therefore we looked at the energy spectrum of final Compton electrons when
increasing the number of driving sub-pulses. Specifically, we investigated the behavior of the probability distributions
of the final electrons:
\begin{equation}
\frac{d^3\mathsf{P}_{\rm C}(p_{\mathrm{f}},\sigma)}{dE_{{\bm p}_{\mathrm{f}}}d^2\Omega_{\mathrm{f}}}
=\frac{1}{2}\sum_{\lambda_{\rm i},\lambda_{\rm f}}\frac{\alpha(m_{\mathrm{e}}c)^2 k^0 |\bm{p}_{\mathrm{f}}|}{(2\pi)^2 E_{{\bm p}_{\mathrm{i}}} (k\cdot K)}|\mathcal{A}_{\mathrm{C}}|^2,
\label{compton5}
\end{equation}
with $\mathcal{A}_{\mathrm{C}}$ defined by Eq.~\eqref{compton2} [see, also the discussion that follows Eq.~\eqref{compton2}].
In Fig.~\ref{boost20130330}, we show the respective distribution in the reference frame in which the initial electron moves in 
the direction of the laser pulse propagation with a momentum $|{\bm p}_{\rm i}|=10^{-6}m_{\rm e}c$. The results concern 
a fixed final electron direction, specified by the angles $\theta_{\mathrm{f}}=0.2\pi$ and $\varphi_{\mathrm{f}}=0$. 
As we see, the exactly same properties of the electron combs are observed 
as for the radiation combs. Namely, the comb momenta with a constant separation are recognized, whose position on the energy scale 
do not change with increasing the number $N_{\rm rep}$. At the same time, the heights of those combs scale like $N_{\rm rep}^2$ 
which is in accordance with the phase calculations, similar to the ones presented in Fig.~\ref{combphase20130317f}.

The above results demonstrate the universality of the comb structures in strong-field QED, as they can also be generated 
for electron beams. This opens new possibilities of exploring relativistic optics of matter waves. Moreover, 
to emphasize the general character of comb structures generated in fundamental processes of strong-field QED, 
we show in Fig. \ref{bwcomb20130322par} the formation of the comb energy spectrum for positrons created via the 
Breit-Wheeler process~\cite{KKbw,breitothers}. The exact same properties 
of the presented spectra as those discussed above are observed.

To conclude, we have demonstrated in this Letter the formation of coherent comb structures in fundamental processes 
of strong-field QED, that for the Compton scattering can be experimentally verified with presently available laser 
and electron beam parameters. These structures relate to both radiation and particle spectra, having the potential to
impact a broad range of scientific studies. In particular, the radiation combs can be used for the generation of zepto- 
and even yoctosecond laser pulses. It appears that subtle characteristics of these structures, like the distance between 
the peaks, can be further controlled by delaying sub-pulses with respect to each other. Such problems 
are going to be presented in future publications.

This work is supported by the Polish National Science Center (NCN) under Grant No. 2011/01/B/ST2/00381.

\end{document}